\documentclass[final,5p,times,twocolumn]{elsarticle}

\usepackage{amssymb}
\usepackage{amsmath}
\usepackage{threeparttable}
\usepackage{float}
\usepackage{rotating}  
\biboptions{sort&compress}
\setcitestyle{square}
\usepackage{url}
\usepackage[ruled,vlined,linesnumbered]{algorithm2e}
\usepackage{hyperref}
\hypersetup{
    colorlinks=true,
    linkcolor=blue,
    filecolor=magenta,      
    urlcolor=cyan,
    pdftitle={Overleaf Example},
    pdfpagemode=FullScreen,
    }
\begin{document}

\begin{frontmatter}



\title{Peak finding algorithm for cluster counting with domain adaptation}


\author[ihep,ucas]{Guang Zhao \corref{cor1}}
\ead{zhaog@ihep.ac.cn}
\author[ihep,ucas]{Linghui Wu}
\author[INFN_Lecce]{Francesco Grancagnolo}
\author[INFN_Bari,U_Bari]{Nicola De Filippis}
\author[ihep,ucas]{Mingyi Dong}
\author[ihep,ucas]{Shengsen Sun}
\affiliation[ihep]{
            organization={Institute of High Energy Physics},
            addressline={19B Yuquan Rd}, 
            city={Beijing},
            postcode={100049}, 
            country={China}}
\affiliation[ucas]{
            organization={University of Chinese Academy of Science},
            addressline={No.1 Yanqihu East Rd}, 
            city={Beijing},
            postcode={101408}, 
            country={China}}
\affiliation[INFN_Lecce]{
            organization={Istituto Nazionale di Fisica Nucleare},
            addressline={Via Arnesano, snc},
            city={Lecce LE},
            postcode={73100},
            country={Italy}}
\affiliation[INFN_Bari]{
            organization={Istituto Nazionale di Fisica Nucleare},
            addressline={Via Giovanni Amendola, 173},
            city={Bari BA},
            postcode={70126},
            country={Italy}}
\affiliation[U_Bari]{
            organization={Politecnico di Bari},
            addressline={Via Edoardo Orabona, 4},
            city={Bari BA},
            postcode={70126},
            country={Italy}}
\cortext[cor1]{Corresponding author}

\begin{abstract}
Cluster counting in drift chamber is the most promising breakthrough in particle identification (PID) technique in particle physics experiment. Reconstruction algorithm is one of the key challenges in cluster counting. In this paper, a semi-supervised domain adaptation (DA) algorithm is developed and applied on the peak finding problem in cluster counting. The algorithm uses optimal transport (OT), which provides geometric metric between distributions, to align the samples between the source (simulation) and target (data) samples, and performs semi-supervised learning with the samples in target domain that are partially labeled with the continuous wavelet transform (CWT) algorithm. The model is validated by the pseudo data with labels, which achieves performance close to the fully supervised model. When applying the algorithm on real experimental data, taken at CERN with a 180 GeV/c muon beam, it shows better classification power than the traditional derivative-based algorithm, and the performance is stable for experimental data samples across varying track lengths.
\end{abstract}




\end{frontmatter}

%\tableofcontents

%\linenumbers


\section{Introduction}
\label{sec:intro}
In the realm of particle physics, future e$^+$e$^-$ colliders play a crucial role in extending the search for new phenomena, potentially addressing questions that the Standard Model (SM) currently cannot explain. Among the major physics programs, the flavor physics program demands particle identification (PID) performance that surpasses that of the most detectors designed for the current generation. Cluster counting, a technique that quantifies the number of primary ionizations \cite{davidenko1969measurement, cataldi1997cluster} rather than the energy loss (d$E$/d$x$) \cite{blum2008particle} along a particle's trajectory in a gaseous detector, represents a promising breakthrough in PID. The Poissonian nature of cluster counting provides a statistically robust method for ionization measurement, potentially yielding a resolution twice as precise as d$E$/d$x$ \cite{perrino2009cluster}. Drift chambers (DC) equipped with cluster counting have been proposed as advanced detector candidates for both the Circular Electron Position Collider (CEPC) \cite{cepc2018cepc} and the Future Circular Collider (FCC) \cite{Gaudio:2022jve}. These detectors are essential tools for unraveling the mysteries of particle physics, offering insights into the properties and interactions of fundamental particles.

\begin{figure}[!htb]
    \centering
    \includegraphics[width=0.5\textwidth]{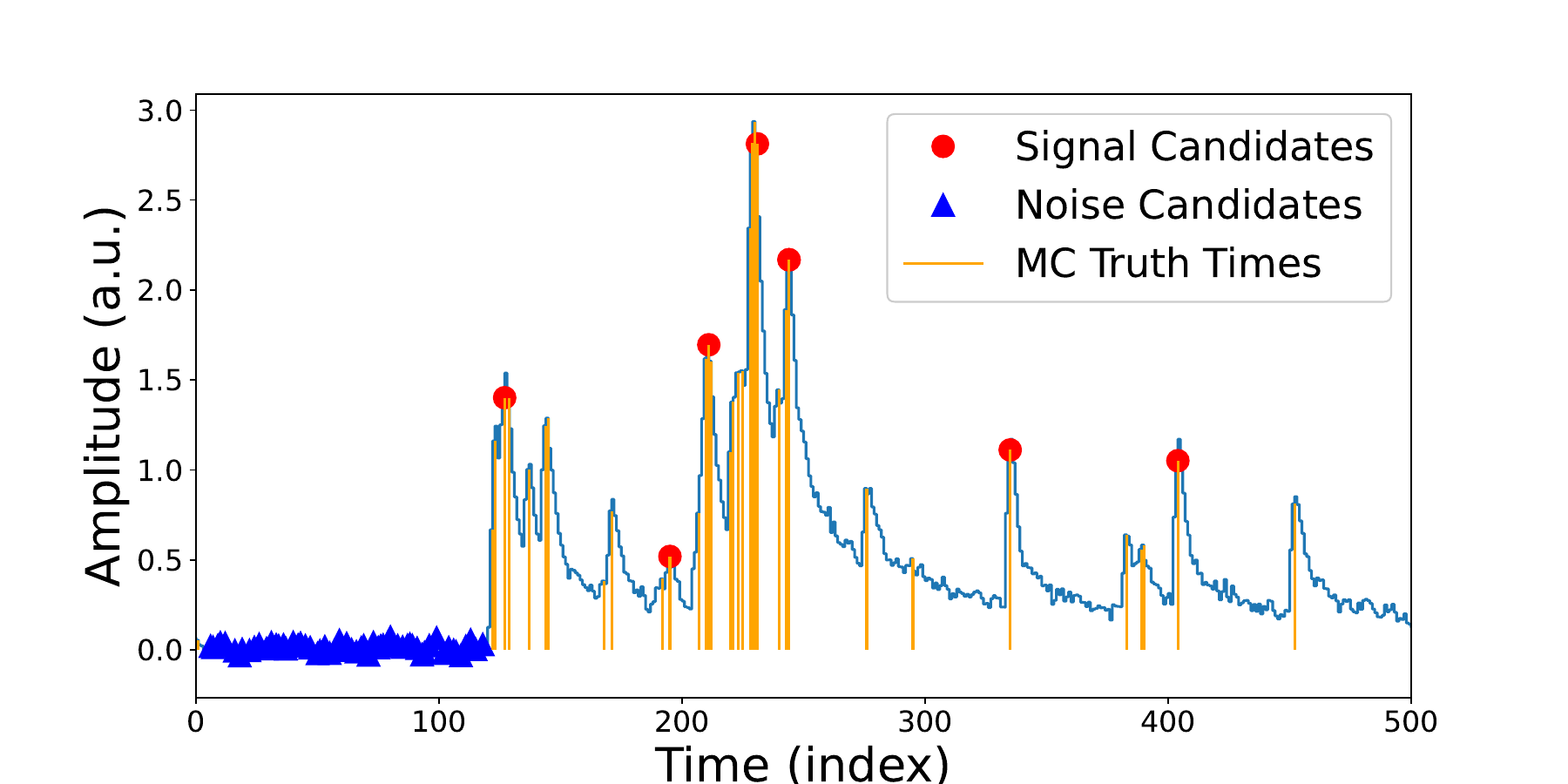}
    \caption{An example of simulated waveform. The blue histogram is the waveform. The red solid circles are the signal peaks selected by the CWT algorithm. The blue solid triangles are the noise peaks selected by requiring the 3 RMS requirement. The orange lines indicate the electron signal times from MC truth information.}
    \label{fig:mc_wf}
\end{figure}

One major challenge of the cluster counting is the efficient reconstruction algorithm to recover the clusters signals (in the form of peaks) in the waveform from a DC measurement. These cluster signals tend to be stacked together in noisy environments (Fig. \ref{fig:mc_wf}). Traditional methods often fall short of achieving the necessary efficiency due to sub-optimal information utilization. For instance, in the derivative-based method (section \ref{sec:derivative}), the peaks are detected by analyzing derivatives and imposing threshold criteria. However, only the rising edges of the single pulses, instead of the full pulse shape, are effectively used. The machine learning (ML) algorithms, which are designed to harness large datasets to reduce complexity and find new features in data, are the state-of-the-art in PID.

Traditional ML relies on fully labeled samples for supervised learning. Such method is well-suited for Monte Carlo (MC) simulated samples, where the MC samples are generated through first-principle simulation, ensuring perfect labels. However, when applying ML algorithm to real experimental data samples, it is usually difficult to apply fully supervised learning. Usually more or less significant discrepancies exist between the simulation and the experimental data, which leads to a degraded performance of the supervised model trained solely on simulated sample. In some cases, these models may even fail to predict experimental data accurately. On the other hand, the experimental data often lack precise labels for fully supervised learning.

Domain adaptation (DA) aims at alleviating this issue by transferring knowledge between domains \cite{pan2009survey}. In our context, the MC simulated sample serves as the source domain, while the experimental data represents target domain. A principled approach to solving domain adaptation involves aligning the source domain distribution with that of the target domain. This alignment allows us to leverage labeled data from the source domain to train a classifier in the aligned domain, which can then be applied to the target domain. In reference \cite{flamary2016optimal}, the Optimal Transport (OT) is exploited to align the source and target domains. One important property of OT is the geometry sensitivity, which facilitates distance computation between probability distributions. By relaxing the invariant requirement on conditional distribution, reference \cite{courty2017joint} proposed the Joint Distribution Optimal Transport (JDOT) , in which both marginal feature and conditional distributions are adapted by minimizing a global divergence between them. A further improvement on the path is to integrate the deep neural network and enable large scale learning \cite{damodaran2018deepjdot}. 

In the context of these studies, models are trained with a target sample without any labels, which is also known as the unsupervised domain adaptation. In our research, we build an approach upon the work presented in reference \cite{damodaran2018deepjdot} and create a semi-supervised version of the domain adaptation model. The model will be employed for analyzing data collected during the drift-tube-prototype experiment conducted at CERN \cite{cern}. We make our code available at \url{https://github.com/littlepi/PeakFindingSemiDA}.

\section{Data Samples for Cluster Counting}
\label{sec:dataset}

The major mechanism underlying for particle identification in gaseous detector is the ionization of matter by charged particles. For DC, each electron produced through ionization generates a distinct peak in the readout waveform. If the electronics is sufficiently fast, these peaks become quantifiable. The reconstruction algorithm plays a crucial role in identifying and associating these peaks with individual electrons. The waveforms can be collected either from test beam experiment or from simulations. In the subsequent sections, we delve into the specifics of the collected data samples.

\subsection{Test beam experiment}
\label{sec:data_sample}
\begin{figure*}[!ht]
    \centering
    \rotatebox{270}{\includegraphics[width=0.4\textwidth]{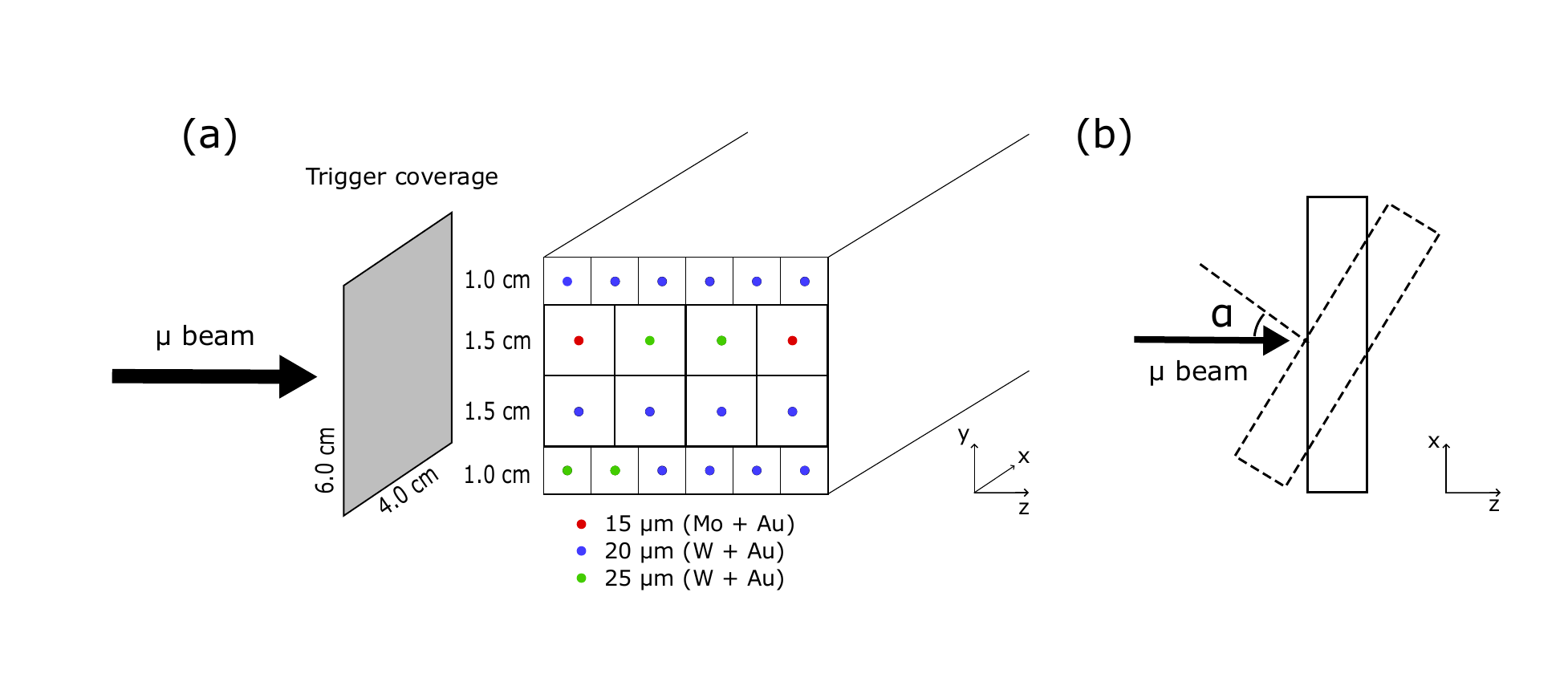}}
    \caption{Schematic representation of the 2022 test beam experiment setup at CERN. (a) Front view. (b) Top view.}
    \label{fig:test_beam}
\end{figure*}

Since 2021, a series of three test beam experiments have been conducted at the CERN/H8 beam line \cite{caputo2023particle}. These experiments employ muon beams with momenta ranging from 4 GeV/c to 180 GeV/c. The experimental setup consists of drift tubes, with dimensions varying between 1 and 3 cm. These tubes are instrumented in collecting data using gas mixtures, specifically He/iC$_{4}$H$_{10}$ in ratios of 90/10 and 80/20. The readout electronics employed in these experiments boasts impressive specifications: a 1 GHz bandwidth and a sampling rate of 1 GS/s at 12 bits. This high-speed electronics configuration ensures excellent single pulse shape resolution, which is crucial for accurate cluster counting. Figure \ref{fig:test_beam} shows a schematic representation of the experimental setup for the 2022 test beam. The resulting waveforms are initially stored in binary files, and subsequently converted into ROOT files \cite{root}.

\subsection{MC Simulation}
\label{sec:mc_sample}
\begin{figure*}[!ht]
    \centering
    \includegraphics[width=1.0\textwidth]{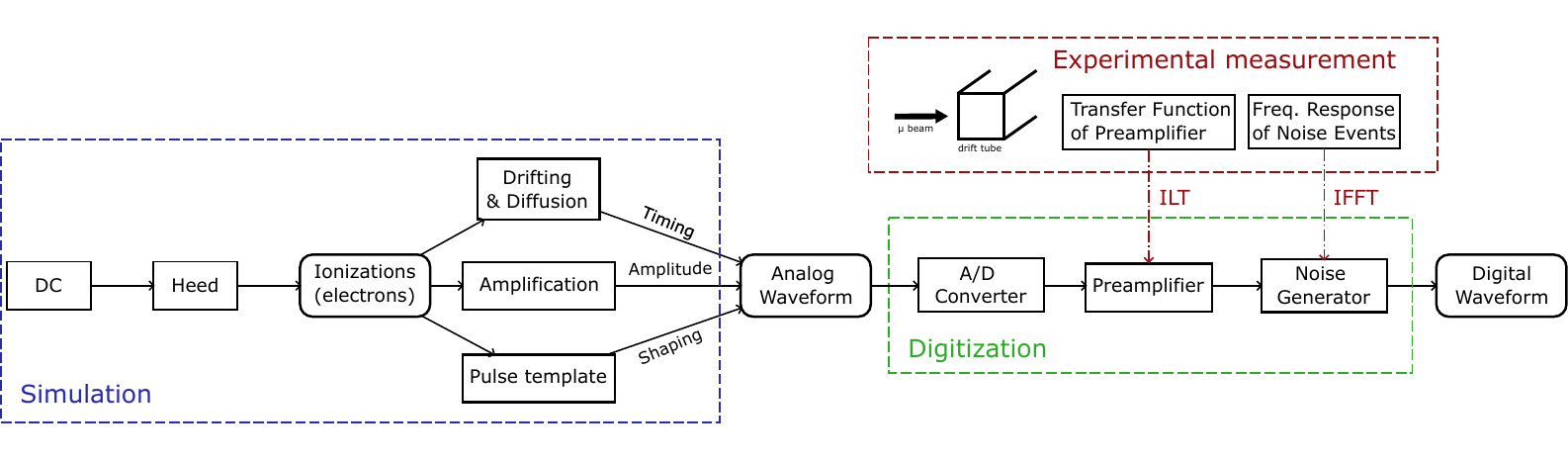}
    \caption{Overview of the MC simulation package for cluster counting. The simulation package creates analog induced current waveforms from ionizations. The digitization package incorporates electronics responses taken from experimental measurements and generates realistic digital waveforms.}
    \label{fig:simulation}
\end{figure*}

A sophisticated waveform-based simulation framework has been developed to generate realistic waveforms for cluster counting (Fig. \ref{fig:simulation}). This framework consists of two essential components: simulation and digitization. 

In the simulation, Garfield++-based code with parameterizations has been developed \cite{garfield}. In the Garfield++ simulation, parameters such as detector geometry, gas mixture and high voltage are configured to match the test beam experiment setup. The ionization patterns resulting from relativistic charged particles are simulated using the Heed \cite{smirnov2005modeling}. To mitigate computational expenses, the following amplification and induced current waveform generation (typically resource-intensive in Garfield++) have been replaced with parameterization methods. 
The timing, amplitude and shape information of the single pulse is parameterized according to Garfield++ full simulations. The parameterized single pulses from ionizations are assembled to an analog induced current waveform from a DC cell.

The generated analog waveforms are further digitized to incorporate realistic electronics responses. The impulse response of the pre-amplifier in time domain is calculated by taking inverse Laplace transform (ILT) on the transfer function in $s$-domain, which is measured from experiment. It is then convoluted with the induced current waveform. Frequency response of the electronics noises is obtained by performing Fourier analysis on the noise waveforms from the experimental data. The noises in the time domain are recovered by taking inverse fast Fourier transform (IFFT) while assuming random phases on the frequency response and added to the waveform. The digitization outputs digital waveforms with data-driven electronics responses.

Figure \ref{fig:mc_wf} illustrates a typical simulated waveform, where the orange lines represent the electron signal positions obtained from the MC truth information. Notably, the waveform exhibits several distinct peaks corresponding to electron signals, often with multiple piled-up signals in significant pulses. It is worth mentioning that locating electron signals near the valleys of the waveform is relatively rare.

\subsection{Data Preprocessing}
\label{sec:pre_processing}
The task of peak finding can be framed as a classification problem in machine learning. The waveforms are divided into segments, each comprising 15 bins. Each segment can represent either a signal or a noise. The list of the amplitudes of a segment, subtracted by their mean and normalized by their standard deviation, is served as the input feature for the neural network. For simulated waveforms, full labels can be obtained from the first-principle MC truth information. For waveforms from experimental data, exact labels may not be available. However, traditional algorithms can still infer some labels. An algorithm based on Continuous Wavelet Transformation (CWT) (section \ref{sec:cwt}) is employed to select signal candidates. Strict criteria ensure high-confidence selection of these candidates. For noise candidates, waveform segments whose maximum amplitudes are smaller than three times the root mean square (RMS) of the noise waveforms are chosen. Figure \ref{fig:mc_wf} illustrates the artificially selected signal and noise candidates with high confidence level.

\section{Semi-supervised Domain Adaptation Model}
\label{sec:model}

%\subsection{Optimal Transport and Deep Joint Distribution Optimal Transport}
%\label{sec:deepJDOT}
Optimal transport theory can be informally described using the words of the French mathematician Gaspard Monge \cite{monge1781memoire}: A worker with a shovel in hand has to move a large pile of sand lying on a construction site, while minimizing the total effort, quantified for instance as the total distance or time spent carrying shovelfuls of sand. The original Monge problem is non-convex, and is difficult to solve \cite{peyre2017computational}. Addressing to this issue, Kantorovich makes the relaxation by moving away the deterministic nature of transportation \cite{kantorovich1942translocation}. The discrete version of the Kantorovich formalism is 
\begin{equation}
\label{eq:ot}
    \gamma_{0}=\underset{\gamma \in \mathcal{P}}{\arg\min} \left\{ \left\langle\gamma,C\right\rangle_{F}=\sum_{i,j}\gamma_{i,j}c_{i,j} \right\},
\end{equation}
where $C$ is a cost matrix, $\gamma$ is a probabilistic coupling and $\mathcal{P}$ is the marginal constraint. Optimal transport has a rich history of application. Thanks to the emergence of approximate solvers that can scale to large problem dimensions, OT is widely used in imaging sciences, graphics or machine learning in recent years \cite{peyre2019computational, bonneel2023survey, torres2021survey}.

Damodaran et al. \cite{damodaran2018deepjdot} proposed the unsupervised domain adaptation model with OT. The unsupervised domain adaptation is a learning framework to transfer knowledge learned from source domains with a large number of annotated training examples to target domains with unlabeled data only. Optimal transport plays an important role in aligning the joint feature and discriminate information between the source and target samples. The objective function in reference \cite{damodaran2018deepjdot} can be expressed as 
\begin{equation}
\label{eq:unsup_ot}
\begin{split}
    \min_{\gamma, f, g}&\frac{1}{n_s}\sum_{i}L_{s}\left(y_i^s,f(g(x_i^s))\right) + \\
    & \sum_{i,j}\gamma_{ij}\left(\alpha\parallel g(x_j^s)-g(x_j^t) \parallel^2 + \lambda_t L_t (y_i^s,f(g(x_j^t)))\right),
\end{split}
\end{equation}
where $x$ is the input feature, $y$ is the label, the superscript $s(t)$ indicates samples from the source (target), and the subscript denotes the $i(j)$-th sample in the source (target); The objective function minimizes over the coupling matrix $\gamma$ of OT, the classifier $f$ and the embedding function $g$; The first term denotes the loss of labeled samples, and the second term denotes OT loss in which the cost function of OT is a combination of differences of features and classifier predictions in the embedding space between the source and target samples; The $1/n_s$, $\alpha$ and $\lambda_t$ are the coefficients for different components.

%\subsection{Semi-supervised Domain Adaptation with Optimal Transport}
For peak finding in cluster counting, as described in section \ref{sec:pre_processing}, partial labels with high confidence levels can be incorporated using the CWT algorithm. Such information can be leveraged for semi-supervised training to enhance the performance of the classifier. Furthermore, a term representing the loss of the labeled samples in target domain is added to the objective function:
\begin{equation}
\label{eq:semisup_ot}
\begin{split}
    \min_{\gamma, f, g}&\frac{1}{n_s}\sum_{i}L_{s}\left(y_i^s,f(g(x_i^s))\right) + \\ 
    & \sum_{i,j}\gamma_{ij}\left(\alpha\parallel g(x_j^s)-g(x_j^t) \parallel^2 + \lambda_t L_t (y_i^s,f(g(x_j^t)))\right) + \\
    & \frac{1}{n_l}\sum_kL_t\left(y_k^{t,l},f(g(x_k^{t,l}))\right).
\end{split}
\end{equation}

In our implementation of the model, the loss function $L_{s(t)}$ is the binary cross entropy. The embedding function $g$ and the classifier $f$ are both fully connected neural networks. For the OT solver, the Earth Movers Distance transport plan is computed using the Python Optimal Transport package \cite{flamary2021pot}. The model is trained iteratively with mini-batches, where in each iteration, the $\gamma$ and ($g$, $f$) is optimized individually while keeping the other fixed.

\section{Numerical Experiments}
\label{sec:exp}
Two numerical experiments are conducted for the peak finding problem. Initially, the semi-supervised DA model is validated with a pseudo data experiment, followed by its application to test beam data samples.

\subsection{Experiment with Pseudo Data}
\label{sec:exp_pdata}
\begin{figure}[!htb]
    \centering
    \includegraphics[width=0.5\textwidth]{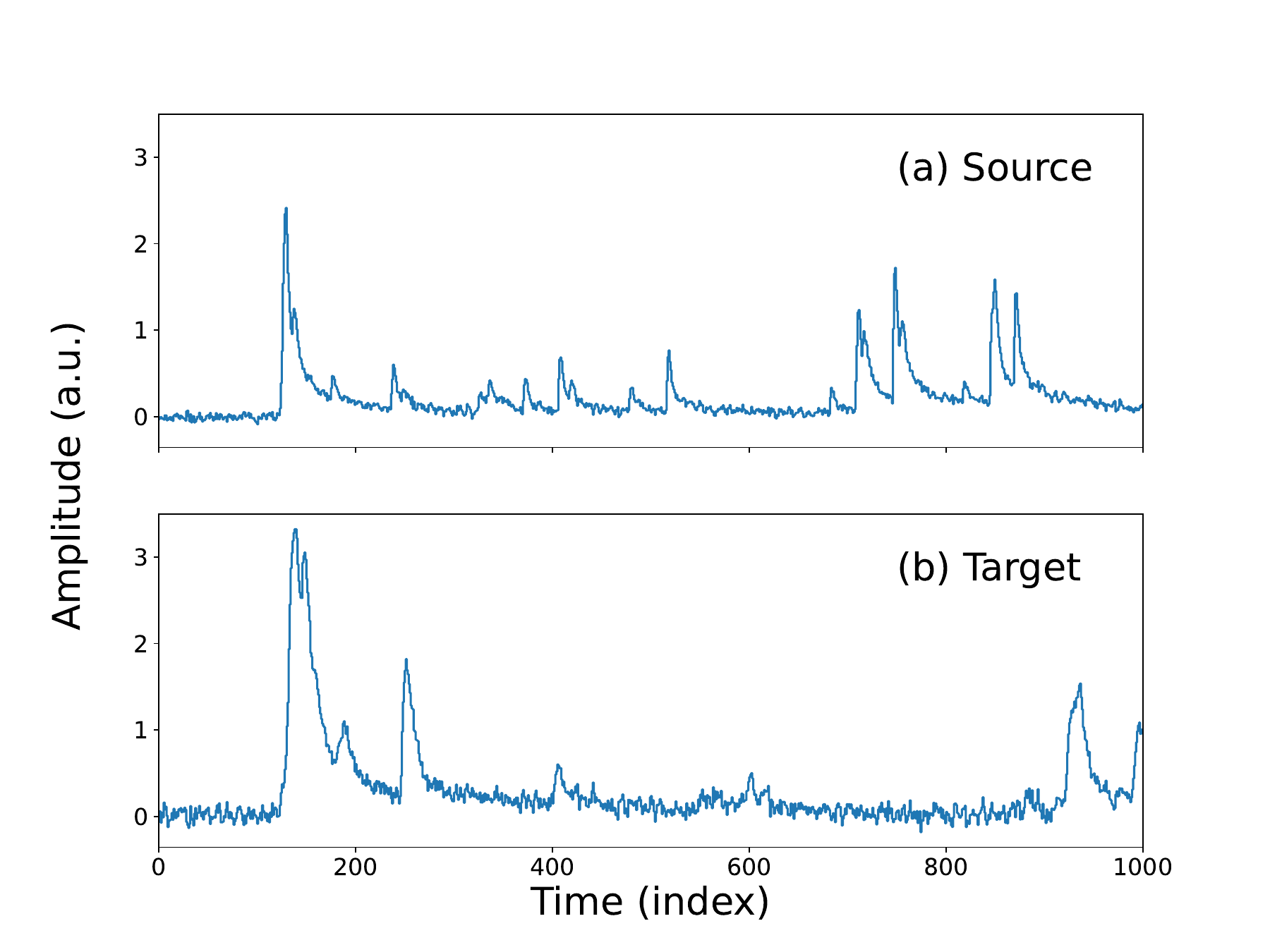}
    \caption{Waveform examples from the source sample (a) and the target sample (b). The source waveforms are generated with a noise level of 10\% and a pulse risetime of 2~ns, while the target waveforms with a noise level of 20\% and a pulse risetime of 4~ns.}
    \label{fig:pseudo_wf}
\end{figure}

In order to quantitatively evaluate the model, two pseudo data samples are generated with the MC simulation: one source sample and one target sample. The source and target samples are generated with varying noise levels and pre-amplifier responses, drawing an analogy between MC sample and the actual data sample in cluster counting study. Notably, the target sample exhibits a higher noise level and slower electronics response. Figure \ref{fig:pseudo_wf} illustrates examples of the generated waveforms, revealing clear discrepancy between the source and the target domains. 

In this experiment, four models are evaluated: the ideal model, the baseline model, the unsupervised DA model and the semi-supervised DA model. The ideal model assumes that full label information is available during training, enabling full supervised learning. It is expected to achieve the best overall performance due to the efficient use of information. The other three models are more realistic and do not rely on labels from the target domain sample. Specifically, the baseline model is trained using the source domain sample and directly applied to the target domain sample. The unsupervised DA model follows the description in Eq. \ref{eq:unsup_ot}. The semi-supervised DA model, developed in this work, is described by Eq. \ref{eq:semisup_ot}.

\begin{figure}[!tbh]
    \centering
    \includegraphics[width=0.55\textwidth]{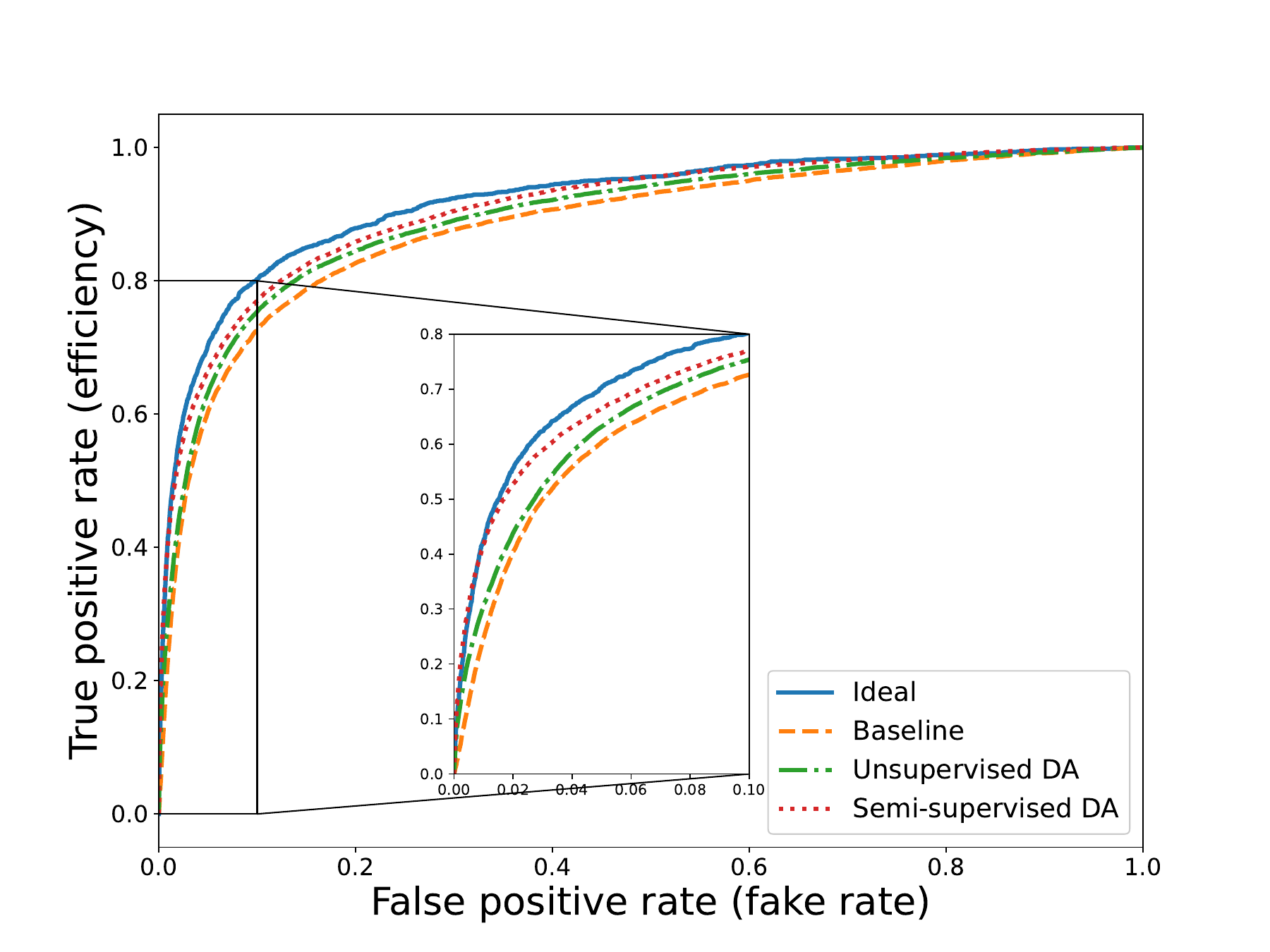}
    \caption{Receiver Operating Characteristic curve for models. The blue curve is from ideal model, the orange curve is from baseline model, the green curve is from unsupervised DA model, and the red curve is from the semi-supervised DA model.}
    \label{fig:roc}
\end{figure}

The labels in the simulated pseudo data samples serve as the basis for evaluating model performance. Figure \ref{fig:roc} illustrates the receiver operating characteristic (ROC) curve, and Tab. \ref{tab:auc} provides the area under curve (AUC) values for various models. As anticipated, the ideal model demonstrates superior performance. In contrast, the baseline model performs poorly among the realistic models due to discrepancies between the source and target domains. The unsupervised DA model, incorporating OT, demonstrates improved performance, indicating that geometric metric alignment is effective for peak finding. Furthermore, the semi-supervised DA model, through semi-supervised learning, achieves even better results overall. Notably, the improvement is most pronounced in the region with a small false positive rate (FPR), which is critical for cluster counting distinguishing particles of different species.

\begin{table}[!htb]
    \centering
    \begin{threeparttable}
    \caption{Area under curve (AUC) and partial-AUC results for models. The partial-AUC's are the normalized AUC values with false positive rates (FPR) smaller than 0.1.}
    \begin{tabular}{l|c|c}
        \hline
        \hline
        Model & AUC & pAUC (FPR$<0.1$) \\
        \hline
        Ideal & 0.926 & 0.812 \\
        Baseline & 0.878 & 0.749 \\
        Unsupervised DA & 0.895 & 0.768 \\
        \textbf{Semi-supervised DA} & \textbf{0.912} & \textbf{0.793} \\
        \hline
        \hline
    \end{tabular}
    \end{threeparttable}
    \label{tab:auc}
\end{table}

\subsection{Experiment with Test Beam Data}
\label{sec:exp_data}
\begin{figure}[!tb]
    \centering
    \includegraphics[width=0.5\textwidth]{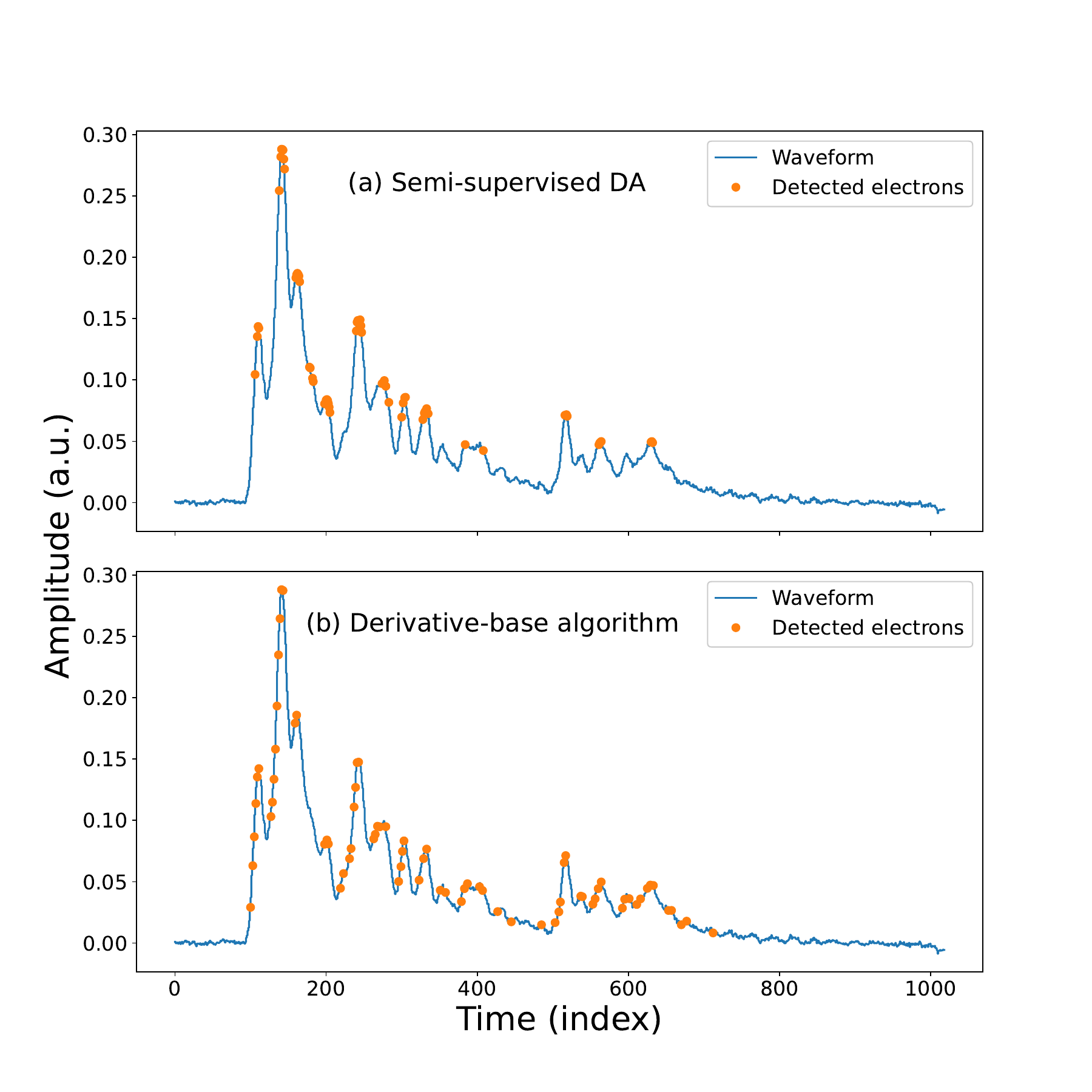}
    \caption{Peak finding on a waveform from test beam data for (a) semi-supervised DA algorithm and (b) derivative-based algorithm. The blue histogram is the original waveform. The orange points are the detected peaks.}
    \label{fig:peak_finding_cmp}
\end{figure}

After validation with the pseudo data sample, the semi-supervised DA model is applied to the real experimental data samples. During model training, the MC simulated sample serves as source domain, while the experimental data sample acts as target domain. Figure \ref{fig:peak_finding_cmp} illustrates an example of applying both the traditional derivative-based algorithm and the semi-supervised DA algorithm to the same waveform from the test beam experiment, while keeping the similar peak finding efficiency. Specifically, when using the derivative-based algorithm, the detected peaks are more dispersed, leading to mis-identification, especially for peaks near the valley. In contrast, the semi-supervised DA algorithm identifies peaks that are mostly clustered around the local maxima, aligning with the expectations from MC simulations (Fig. \ref{fig:mc_wf}). 

\begin{figure*}[!tb]
    \centering
    \includegraphics[width=0.6\textwidth]{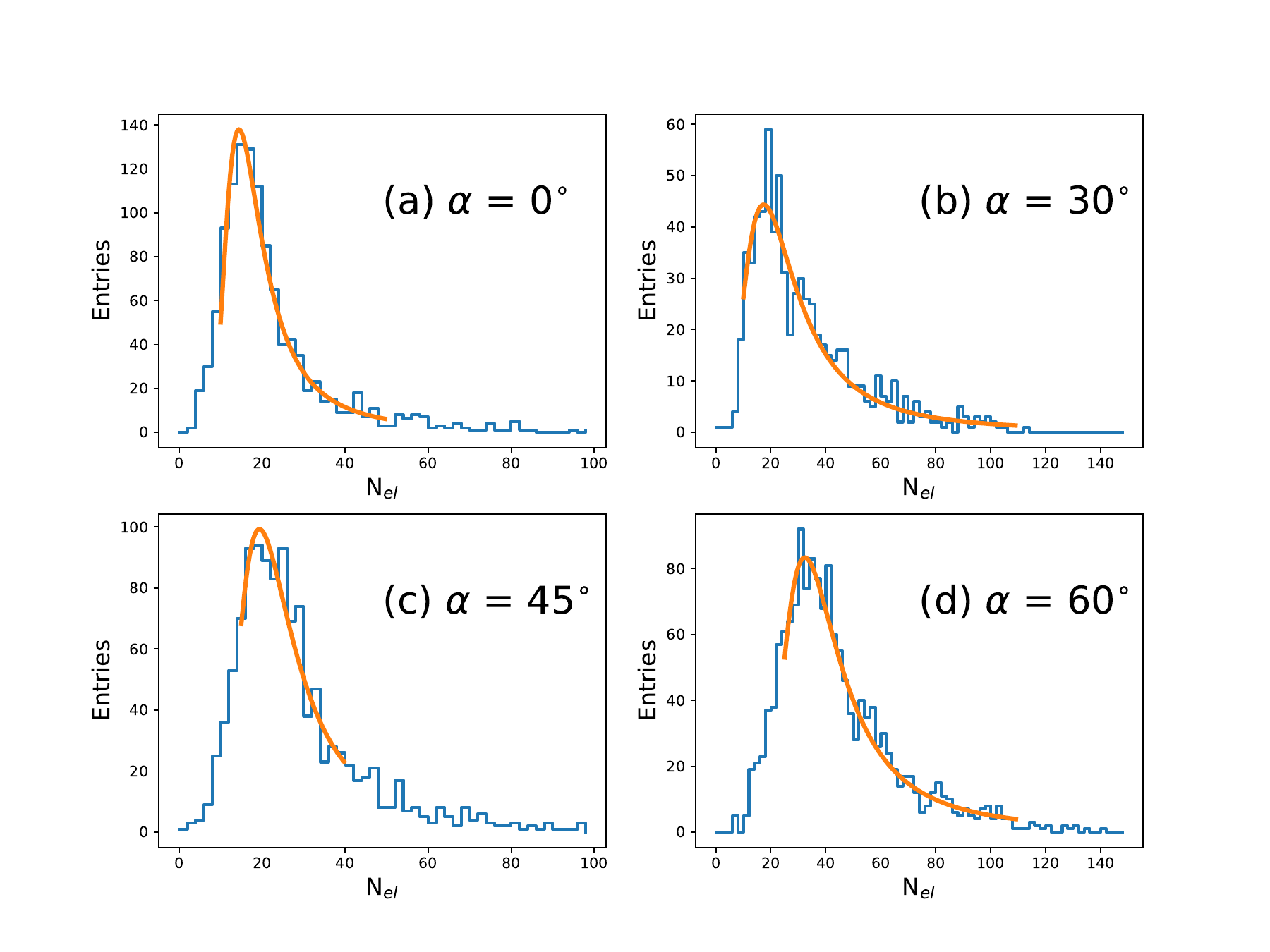}
    \caption{Distribution of number of detected electron signals for samples in $\alpha=0^{\circ}, 30^{\circ}, 45^{\circ}$ and $60^{\circ}$. The orange lines are the Landau-fit curves.}
    \label{fig:ne}
\end{figure*}

\begin{figure}[!tb]
    \centering
    \includegraphics[width=0.5\textwidth]{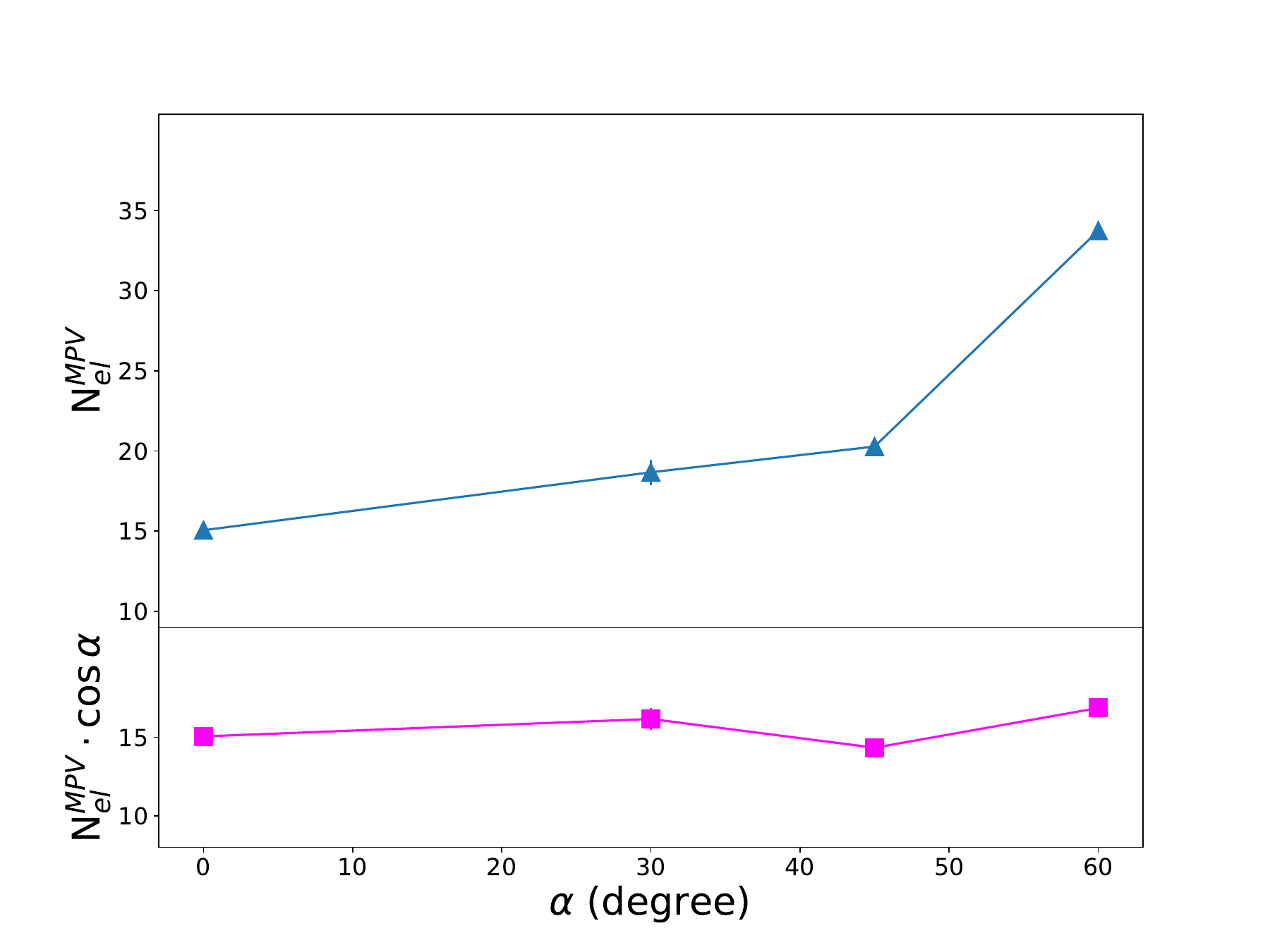}
    \caption{Most probable values of distributions of detected peaks as a function of angles. The blue triangles are the original results ($N_{el}^{MPV}$). The magenta squares are the normalized results w.r.t. the track lengths ($N_{el}^{MPV}\cdot\cos\alpha$).}
    \label{fig:angle}
\end{figure}

The model is applied to additional data samples with varying angles ($\alpha$) between the muon track and the normal of sense wire (Fig. \ref{fig:test_beam} (b)). The distributions of the number of detected peaks ($N_{el}$) for these samples are depicted in Fig. \ref{fig:ne}. Remarkably, all these distributions exhibit a well-Landau shape, which validates the rationality of our peak finding algorithm. The most probable values (MPV) of the $N_{el}$'s along with their normalization relative to the track length are presented in Fig. \ref{fig:angle}, which shows good consistency across all samples. This indicates that our algorithm not only outperforms the derivative-based algorithm but also remains stable across varying track lengths.

\section{Summary}
\label{sec:summary}
In this paper, we develop a semi-supervised DA algorithm specifically for the peak finding problem in cluster counting. The algorithm leverages OT, which provides a geometric metric between distributions, to align the samples from the source domain (MC simulations) and the target domain (experimental data). Additionally, our approach incorporates semi-supervised learning using samples from the target domain that are partially labeled with the CWT algorithm. We validate our model using pseudo data with known labels, achieving performance comparable to a fully supervised model. When applied to real experimental data collected at CERN with a 180 GeV/c muon beam, our algorithm demonstrates superior classification power compared to the traditional derivative-based approach. Furthermore, its performance remains stable across experimental data samples with varying angles. 

Cluster counting in drift chamber represents a significant advancement in PID techniques for particle physics experiments. The reconstruction algorithm is a critical challenge in this context. Our semi-supervised DA algorithm achieves state-of-the-art performance for peak finding in real experimental data. Importantly, the algorithm's principles can be extended to other DA problems that exhibit geometric sensitivity. Looking ahead, as more test beam data become available across a wider range of beam momenta, our algorithm will play a crucial role in data analysis and may be implemented in FPGA systems for feasibility studies related to cluster counting.

\section*{Acknowledgements}
%We acknowledge the contribution of our colleagues from ... , for installing and maintaining the experimental setup and for the data collection and preprocessing. We thank CERN for providing very efficiently all the necessary facilities.

We would like to express our gratitude to our colleagues from Università del Salento and INFN Sezione di Lecce, Università Degli Studi di Bari, Politecnico di Bari and INFN Sezione di Bari, and Northwestern University (US), for their invaluable assistance in installing and maintaining the experimental setup, as well as for their contribution to the data collection and preprocessing. We would also like to extend our appreciation to CERN for providing us with all the necessary facilities in an efficient manner.

This research was funded by Institute of High Energy Physics (Chinese Academy of Sciences) Innovative Project on Sciences and Technologies under Contracts Nos. E3545BU210 and E25456U210, National Natural Science Foundation of China (NSFC) under Contract Nos. 12275296.

\appendix

\section{Traditional Algorithms}
\label{sec:tra-alg}

\subsection{Derivative-Based Algorithm}
\label{sec:derivative}
The principle of a derivative-based algorithm is intuitive. If a peak has a fast rising edge, the derivative should be larger than some threshold, and one can detect peaks by requiring threshold passing. The pseudo code of the algorithm is summarized in Algorithm \ref{alg:derivative}. To exclude noise in the baseline, a threshold $T_1$ of approximately three times the RMS of the average noise amplitudes is initially required. Higher order derivatives are efficient for hidden-peak detection, but more sensitive to noise. Therefore, a second-order derivative is applied to the raw waveform, considering the trade-off between efficiency and purity. Then, an integration is applied to the second derivative, and a threshold $T_2$ is required for peak detection. Since large slope of the rising edge is a key feature to characterize the signal,
only the positive derivatives corresponding to it are kept in the calculation.

The derivative-based algorithm is a simple and easy-to-implement algorithm for cluster counting reconstruction. However, it only uses partial information of the waveform and is less efficient in high noise levels. For high noise-level environment, a low-pass digital filter can be applied before applying the derivative-based algorithm. However, this could drastically reduce the efficiency as the signals of clusters are highly piled-up.

\begin{algorithm}[!htb]
\SetAlgoLined
\KwData{Waveform with $n$ bins ($Input[1,...,n]$)}
\KwResult{A list of detected peaks ($Output$)}
\caption{Derivative algorithm}
\label{alg:derivative}
%\SetKwProg{Fn}{Function}{ begin}{end}
%\Fn{func}{
%    code
%}
\Begin{
    $Output \longleftarrow \emptyset$\;
    $D1[1] \longleftarrow 0$\;
    $D2[1] \longleftarrow 0$\;
    $Int[1,...,n] \longleftarrow 0$\;
    \For{$i \longleftarrow 2$ \KwTo $n$} {
        \lIf{$Input[i-1]<T_{1}$}{{\bf continue}}
        $D1[i] \longleftarrow \max\{Input[i]-Input[i-1], 0\}$\;
        $D2[i] \longleftarrow \max\{D1[i]-D1[i-1], 0\}$\;
        \uIf{$D2[i]>0$}{
            $Int[i] \longleftarrow Int[i-1] + D2[i]$\;
        }
        \ElseIf{$Int[i-1]>T_{2}$} {
            $Output \longleftarrow Output \cup \{i-1\}$\;
        }
    }
}
\end{algorithm}

\subsection{Continuous Wavelet Transform Algorithm}
\label{sec:cwt}
The algorithm used in this study is the SciPy implementation \cite{scipy} of the CWT-based algorithm described in reference \cite{pan2009survey}. The algorithm performs a CWT on the raw waveform and obtains a 2D CWT matrix. The peaks in the waveform and the wavelet can form resonances, leading to ``ridge lines'' in the CWT matrix. The algorithm identifies these ridge lines and determines the position and the widths of the peaks by analyzing them. The key parameters of the algorithm include the widths of the wavelets and the minimum signal-to-noise ratio. The default CWT-based algorithm in SciPy can detect peak positions without bias but suffers from low efficiency for pile-ups. The unbiasing feature makes it a good choice for adding signal labels in real exprimental data samples.

%%\bibliographystyle{elsarticle-harv} 
\bibliographystyle{unsrt} 
\bibliography{reference}


\end{document}